\documentclass[letterpaper,iop]{emulateapj}
\pdfoutput=1 
\usepackage{thumbpdf} 
\usepackage{amsmath,amssymb} 
\usepackage{graphicx} 
\usepackage{epsf}

\usepackage{natbib} 
\usepackage[usenames]{color} 
\usepackage{ifpdf}  
\usepackage[T1]{fontenc}

\usepackage{ifxetex}
\ifxetex
  \usepackage{unicode-math}
  \usepackage[protrusion=true,expansion=false]{microtype} 
\else
  \usepackage{mathptmx}
  \ifpdf
    \usepackage[protrusion=true,expansion=true]{microtype} 
  \else
  \fi
\fi
\usepackage{relsize}

\usepackage{xcolor}[2007/01/21] 
\usepackage{refcount} 
\usepackage{textcase} 
\usepackage{epstopdf} 

\ifpdf
\pdfoutput=1 
\else

\fi

\newcommand{\beq}{
\begin{equation}
}
\newcommand{\eeq}{
\end{equation}
}
\newcommand{\beqa}{
\begin{eqnarray}
}
\newcommand{\eeqa}{
\end{eqnarray}
}

\newcommand{\kgfigbeg}[1]{
\begin{figure}
\hypertarget{#1}{}%
}
\newcommand{\kgfigend}[2]{
\label{f:#1}
\end{figure}
\bookmarksetup{color=[rgb]{0.54,0,0}}
\bookmark[rellevel=1,keeplevel,dest=#1]{Fig \ref*{f:#1}: {#2}}
\bookmarksetup{color=black} 
}

\newcommand{\kgfigstarbeg}[1]{
\begin{figure*}
\hypertarget{#1}{}%
}
\newcommand{\kgfigstarend}[2]{
\label{f:#1}
\end{figure*}
\bookmarksetup{color=[rgb]{0.54,0,0}}
\bookmark[rellevel=1,keeplevel,dest=#1]{Fig \ref*{f:#1}: {#2}}
\bookmarksetup{color=black} 
}

\newcommand{\kgtabbeg}[1]{
\begin{deluxetable}{#1}
}
\newcommand{\kgtabend}[2]{
\label{t:#1}
\end{deluxetable}
\bookmarksetup{color=[rgb]{0,0,0.54}}
\bookmark[
rellevel=1,
keeplevel,
dest=table.\getrefnumber{t:#1}
]{Table \ref*{t:#1}: #2}
\bookmarksetup{color=[rgb]{0,0,0}}
}

\newcommand{\kgtabstarbeg}[1]{
\begin{deluxetable*}{#1}
}
\newcommand{\kgtabstarend}[2]{
\label{t:#1}
\end{deluxetable*}
\bookmarksetup{color=[rgb]{0,0,0.54}}
\bookmark[
rellevel=1,
keeplevel,
dest=table.\getrefnumber{t:#1}
]{Table \ref*{t:#1}: #2}
\bookmarksetup{color=[rgb]{0,0,0}}
}

\newcommand{\units}[1]  {\ensuremath{\mathrm{{#1}}}}
\newcommand{\msun}     {\ensuremath{{{M}}_{\scriptscriptstyle \odot}}}

\newcommand{\mbh}      {\ensuremath{M}}

\providecommand{\ion}[2]{#1$\;$\textsmaller{\@Roman{#2}}}

\def\spose#1{\hbox to 0pt{#1\hss}}
\newcommand{\lta}{\mathrel{\spose{\lower 3pt\hbox{$\mathchar"218$}}
      \raise 2.0pt\hbox{$\mathchar"13C$}}}
\newcommand{\gta}{\mathrel{\spose{\lower 3pt\hbox{$\mathchar"218$}}
      \raise 2.0pt\hbox{$\mathchar"13E$}}}
\def\simlt{\mathrel{\rlap{\lower 3pt\hbox{$\sim$}}\raise 2.0pt\hbox{$<$}}}
\def\simgt{\mathrel{\rlap{\lower 3pt\hbox{$\sim$}} \raise 2.0pt\hbox{$>$}}}

\definecolor{KayhanCiteColor}{rgb}{0,0.08,0.35}
\definecolor{KayhanURLColor}{rgb}{0,0.08,0.35}
\definecolor{KayhanLinkColor}{rgb}{0,0.08,0.35}
\definecolor{KayhanPageColor}{rgb}{0,0.08,0.35}
\definecolor{medred}{rgb}{0.75,0.0,0.0}

\shorttitle{Low-Mass AGN on the Fundamental Plane}
\shortauthors{G\"{u}ltekin et al.}

\submitted{Accepted by The Astrophysical Journal Letters}

\makeatletter
\ifx\undefined\hyperref
\usepackage[pdftitle={Low-Mass AGN and Their Relation to the Fundamental Plane of Black Hole Accretion},pdfauthor={Kayhan Gultekin},pdfsubject={Astrophysics},pdfkeywords={black hole physics --- galaxies: general --- galaxies: nuclei --- galaxies: bulges --- methods: statistical},letterpaper,linktocpage,colorlinks=true,linkcolor={KayhanLinkColor},urlcolor={KayhanURLColor},citecolor={KayhanCiteColor},hyperfootnotes=false]{hyperref}
\else\relax\fi
\makeatother
\usepackage[figure,figure*]{hypcap} 
\usepackage{atveryend} 
\usepackage[
  open,
  openlevel=3,
  atend
]{bookmark}[2010/04/08]  
\begin{document}

\label{firstpage}
 
\title{Low-Mass AGN and Their Relation to the Fundamental Plane of Black Hole Accretion}

\author{Kayhan G\"{u}ltekin\altaffilmark{1}}
\author{Edward M.\ Cackett\altaffilmark{2}}
\author{Ashley L.\ King\altaffilmark{1}}
\author{Jon M.\ Miller\altaffilmark{1}}
\author{Jason Pinkney\altaffilmark{3}}
\affil{\altaffilmark{1}Department of Astronomy, University of Michigan, 500 Church Street, Ann Arbor, MI 48109, USA;
\href{mailto:kayhan@umich.edu}{kayhan@umich.edu}}
\affil{\altaffilmark{2}Department of Physics and Astronomy, Wayne State University, 666 West Hancock Street, Detroit, MI 48201, USA}
\affil{\altaffilmark{3}Department of Physics and Astronomy, Ohio Northern University, 525 S.\ Main St., Ada, OH 45810, USA }

\begin{abstract}
\hypertarget{abstract}{} 
We put active galactic nuclei (AGNs) with low-mass black holes on the
fundamental plane of black hole accretion---the plane that relates
X-ray emission, radio emission, and mass of an accreting black
hole---to test whether or not the relation is universal for both
stellar-mass and supermassive black holes.  We use new \emph{Chandra}
X-ray and Very Large Array radio observations of a sample of black holes with
masses less than $10^{6.3}\ \msun$, which have the best leverage for
determining whether supermassive black holes and stellar-mass black
holes belong on the same plane.  Our results suggest that the two
different classes of black holes both belong on the same relation.  These
results allow us to conclude that the fundamental plane is suitable
for use in estimating supermassive black hole masses smaller than
$\sim 10^7\ \msun$, in testing for intermediate-mass black holes, and
in estimating masses at high accretion rates.  
\bookmark[ rellevel=1, keeplevel,
dest=abstract
]{Abstract}
\end{abstract}

\section{Introduction}
\label{intro}


The continuum X-ray and radio emission and mass of accreting black
holes are empirically correlated \citep{merlonietal03, fkm04}.  The
correlation, frequently called the ``fundamental plane of black hole
accretion,'' relates mass accretion rates, probed by the X-ray
luminosity, and jet or outflow power, probed by the radio luminosity,
at a given mass of the black hole, which sets the size scale of the
accretion-disk--jet system.  The relation was first seen in
stellar-mass black holes, which have a narrow range in mass
\citep{2003MNRAS.344...60G}, and then extended to all black holes
\citep{merlonietal03, fkm04}, covering over 8 orders of magnitude in
mass, 14 orders of magnitude in radio luminosity, and 12 orders of
magnitude in X-ray luminosity.

Despite the wide range that the fundamental plane covers, there is
room to question its universality.  For example, using a larger sample
of stellar-mass black hole data, \citet{2012MNRAS.423..590G} found
that the radio--X-ray correlation was best explained by two tracks,
with one source that transitions from one track to the other.
Restricting to only black holes with dynamical mass measurements,
\citet{2009ApJ...706..404G} found different results when including and
excluding stellar-mass black holes ($L_R\sim M^{0.78}L_X^{0.67}$ and
$L_R\sim M^{2.08}L_X^{0.50}$, respectively).

There are two potential explanations for the separate relations found
by \citet{2009ApJ...706..404G}.  The first is that supermassive black
holes (SMBHs) and stellar-mass black holes do occupy the same fundamental
plane, and the apparent difference is a
result of a small sample size and limited dynamic range.  The second
is that SMBHs and supermassive black holes each
operate under different physical conditions and follow their own
relation, and fits to combined samples only appear to produce a single
relation because of the large logarithmic range in values.  Either
empirical result would provide insight to the physics at play.

In this Letter, we report an observational study designed to test
which empirical relation is better.  We selected a sample of active galactic nuclei (AGNs) with
masses in the range that is best suited to make an observational
distinction between the two relations (Section \ref{sample}).  We
obtained new \emph{Chandra} and Karl G.\ Jansky Very Large Array (VLA)
observations and supplemented these with archival data to put them on the
fundamental plane (Sections \ref{xrayobs} and \ref{radioobs}).  We
find that the single plane is a better predictor of the observed radio
and X-ray fluxes, and we discuss the implications of our results
(Section \ref{discussion}).  Throughout this Letter we assume $H_0=70\ \units{km\ s^{-1}\ Mpc^{-1}}$, $\Omega_M=0.3$, and
$\Omega_\Lambda=0.7$.

\section{Observations}
\label{observations}

\subsection{Sample Selection}
\label{sample}

To test whether an SMBH-only relation or an all-black-holes relation
was better, we assembled a sample of 13 low-mass AGNs from the sample
of \citet{gh07}.  Black hole masses ($\mbh$) come from the virial
(i.e., single-epoch) method using broad H$\alpha$ emission lines and
an assumed relation between the radius of the broad-line region and
continuum luminosity of the AGN, inferred from the broad H$\alpha$
luminosity.  The parent sample was defined to have
$\mbh<10^{6.3}\ \msun$.  Some of the parent sample were flagged with c
by \citet{gh07} to indicate that the broad-line emission may not have
been robustly detected.  The c-sample sources may therefore have less
robust mass estimates.  As only one c-sample source is detected in
both radio and X-ray, our fundamental-plane analysis and conclusions
below are not affected.

Out of 229 low-mass AGNs, we selected 13 for which the X-ray and radio
fluxes could be measured in a reasonable exposure as predicted by the
\citet{2009ApJ...706..404G} fundamental plane.  To determine the
suitability of potential sources for our new X-ray and radio
observations (Sections \ref{xrayobs} and \ref{radioobs}), we used
information from existing optical observations.  For X-ray
detectability, we assumed the bolometric luminosities from
\citet{gh07} and bolometric correction from
\citet{2007MNRAS.381.1235V}.  For VLA detectability, we assumed the
more conservative of the two \citet{2009ApJ...706..404G} fundamental
plane predictions.

We included both sources that have VLA FIRST
\citep{1997ApJ...475..479W} 1.4 GHz detections and those that
do not, so as not to bias ourselves to those brightest in radio
wavelengths.  As discussed below, VLA FIRST detection was not a good
predictor of 8.5 GHz continuum emission.

To reduce scatter related to source variability, we scheduled
\emph{Chandra} and VLA observations contemporaneously.  We also used
archival X-ray data when possible.  Table \ref{t:sample} lists basic
properties---including \citet{gh07} identification number, which we use here---of all low-mass AGNs in our sample.  We
have updated the masses to use the most recent single-epoch H$\alpha$
scaling relations \citep{2005ApJ...630..122G, 2013ApJ...767..149B,
  2013ApJ...775..116R} and to assume a scaling factor of
$\epsilon=4.3/4$ \citep{2013ApJ...773...90G}, which uses the approximation
$V_\mathrm{FWHM}=2\sigma$.  This makes a typical difference of $<0.1$
dex compared to \citet{gh07}, much smaller
than the estimated 0.5 dex systematic uncertainty.

\kgtabbeg{r@{\extracolsep{0pt}}llrrrr}
  \footnotesize
  \tablecaption{Sample of Small AGN}
  \tablewidth{\columnwidth}
  \tablehead{
     \multicolumn{2}{c}{GH ID} &
     \colhead{SDSS} &
     \colhead{$z$} &
     \colhead{$D_L$} &
     \colhead{$\log(M/\msun)$} &
     \colhead{$\Delta T_\mathrm{obs}$}
  }
  \startdata
 47& & J082443.28+295923.5 & 0.025 & 109 & 5.70 &  1890 \\
 69& & J091449.05+085321.1 & 0.140 & 661 & 6.30 &   176 \\
 87& & J101246.49+061604.7 & 0.078 & 353 & 6.22 &    60 \\
 94&$^\mathrm{c}$ & J103234.85+650227.9 & 0.006 &  26 & 5.80 & \dots \\
101&$^\mathrm{c}$ & J105108.81+605957.2 & 0.082 & 373 & 6.27 & \dots \\
106& & J110501.97+594103.6 & 0.034 & 149 & 5.58 &   822 \\
119&$^\mathrm{c}$ & J112637.74+513423.0 & 0.026 & 114 & 6.16 & \dots \\
140&$^\mathrm{c}$ & J121629.13+601823.5 & 0.060 & 269 & 6.18 &    76 \\
146& & J124035.81$-$002919.4 & 0.081 & 368 & 6.35 &  2824 \\
158&$^\mathrm{c}$ & J131659.37+035319.8 & 0.045 & 199 & 5.84 &   226 \\
163&$^\mathrm{c}$ & J132428.24+044629.6 & 0.021 &  91 & 5.81 &   226 \\
174& & J140829.26+562823.5 & 0.133 & 626 & 6.24 &    68 \\
203& & J155909.62+350147.4 & 0.031 & 136 & 6.31 &   991
\enddata

\tablecomments{Basic sample properties, including identification from
  \citet{gh07}, SDSS name, redshift, luminosity distance in Mpc,
  logarithmic black hole mass in solar units, and time in units of
  days between radio and X-ray observations if both exist.  Only
  sources with an entry for $\Delta T_\mathrm{obs}$ have both X-ray
  (new or archival) and radio data for analysis with the fundamental
  plane, including GH 140 and GH 158, which have only upper limits in
  X-ray, and GH 174, which has only an upper limit in radio.  C-sample
  sources (Section \ref{sample}) are identified with a superscript c.}
\kgtabend{sample}{Summary of Small AGN Sample}

\subsection{X-Ray Observations}
\label{xrayobs}

We used five archival and six new \emph{Chandra} observations to measure
X-ray luminosities.  For new observations, exposure times were 15 ksec
except for the dimmest, GH 140, which had an exposure time of 24 ksec.
New \emph{Chandra} observations were observed on the S3 chip of the
ACIS-S detector.  We re-reduced and re-analyzed the archival data.

Data reduction followed the normal pipeline with the most recent
\emph{Chandra} data reduction software package (CIAO version 4.5) and
calibration databases (CALDB version 4.5.5).  Source regions were
circles with radii of 4 or 5 pixels centered on the brightest putative
point source consistent with the center of the host galaxy.  Given the
distances to the host galaxies (Table \ref{t:sample}), X-ray binaries are not 
luminous enough to be a significant source of contamination.
Background regions were annuli with inner radii equal to the
source region radius and outer radii equal to $\sim30$ pixels.  We used the
specextract tool to create response matrix and ancillary
response files and extract source and
background spectra.

Spectral fitting was done with the most recent version of Xspec
\citep[v12.6.0q;][]{1996ASPC..101...17A}.  Since we were most
interested in the unscattered, intrinsic 2--10 keV flux, we always
included a power-law component in our spectral model with Galactic
absorption set to the value toward each source
\citep{2005A&A...440..775K}.  For sources with enough counts, we also
included redshifted intrinsic absorption.  In all cases,
as long as the source was detected and total absorption was below $N_\mathrm{H}=10^{20}\ \units{cm^{-2}}$, the inferred intrinsic 2--10 keV flux
from the power-law component was robust.  For sources with low
count rates or for which only an upper limit could be determined for
the X-ray flux, we held the photon index constant at $\Gamma=1.7$.
Spectral fitting was done with $C$-stat statistics
\citep{1979ApJ...228..939C}.  We report  2--10 keV flux and
$\Gamma$ for each source in Table \ref{t:xrayobs}.  Sources GH 119,
and 140 were nondetections; a point source at the location of GH 158
was detected with more than 12 net counts in the 0.5--10 keV band, but
it was not enough to constrain the hard flux at the $3\sigma$ level.

Source 47 had a 2 ksec archival \emph{Chandra} exposure that showed it
likely a Seyfert 2.  A 23-ksec \emph{XMM-Newton} archival
observation with a high count rate has been looked at in detail by
\citet{reneeinprep}.  They find that the spectrum of GH 47 is well
fitted by a typical type II AGN spectrum consisting of a highly
absorbed power-law component, soft, scattered power-law component, and
a distant reflection component with prominent narrow Fe~K$\alpha$
line.  The unabsorbed 2--10 keV flux of the absorbed power-law
component is $2.05\times 10^{-12}\ \units{erg\ cm^{-2}\ s^{-1}}$.  We
use the \emph{XMM-Newton} measurement for all subsequent analysis.

As a self-consistency check, we calculated Eddington fractions
($f_\mathrm{Edd}$) of each source assuming the bolometric correction
due to \cite{2004MNRAS.351..169M}.  Because the X-ray bolometric
correction depends on the bolometric luminosity, we use the bolometric
luminosity reported in \citet{gh07}.  Although using $f_\mathrm{Edd}$
based on optical emission to estimate $f_\mathrm{Edd}$ based on X-ray
emission is circular, it serves as a check for self-consistency.  Using
a constant bolometric correction of 20 instead of our adopted circular
method does not make a large difference.  We plot the X-ray
$f_\mathrm{Edd}$ as a function of the optically determined
$f_\mathrm{Edd}$ in Figure\ \ref{f:fedd}.  The agreement between the two
estimates shows self-consistency.

\kgtabbeg{rrrrr@{\extracolsep{0pt}}ll}
  \footnotesize
  \tablecaption{X-ray Observations}
  \tablewidth{\columnwidth}
  \tablehead{
     \colhead{GH ID} &
     \colhead{Obsid} &
     \colhead{MJD} &
     \colhead{$t_\mathrm{exp}$} &
     \multicolumn{2}{c}{$\log(F_{X}/\units{erg\,s^{-1}\,cm^{-2}})$} &
     \colhead{$\Gamma$}
  }
  \startdata
 47 & 504102001 & 54408 &  23 &    &$-11.69^{+0.08}_{-0.04}$ & $2.0\pm{0.2}$\\
 69 & \dataset[ADS/Sa.CXO#Obs/13858]{13858} & 56097 & 15 &    &$-12.43\pm{0.03}$ & $1.9\pm{0.1}$\\
 87 & \dataset[ADS/Sa.CXO#Obs/13859]{13859} & 56214 & 15 &    &$-12.63\pm{0.04}$ & $2.0\pm{0.2}$\\
106 & \dataset[ADS/Sa.CXO#Obs/11456]{11456} & 55424 &  2 &    &$-12.17^{+0.05}_{-0.08}$ & $1.6^{+0.2}_{-0.1}$\\
119 &  \dataset[ADS/Sa.CXO#Obs/09234]{9234} & 54551 &  5 & $<$&$-14.39$ & 1.7\\
140 & \dataset[ADS/Sa.CXO#Obs/13860]{13860} & 56200 & 24 & $<$&$-14.05$ & 1.7\\
146 &  \dataset[ADS/Sa.CXO#Obs/5664]{5664} & 53428 &  5 &    &$-12.94^{+0.14}_{-0.09}$ & 1.7\\
158 & \dataset[ADS/Sa.CXO#Obs/13861]{13861} & 56050 & 15 & $<$&$-12.88$ & 1.7\\
163 & \dataset[ADS/Sa.CXO#Obs/13862]{13862} & 56050 & 15 &    &$-12.80\pm{0.07}$ & $1.4\pm{0.2}$\\
174 & \dataset[ADS/Sa.CXO#Obs/13863]{13863} & 56208 & 15 &    &$-12.58\pm{0.03}$ & $1.8\pm{0.1}$\\
203 & \dataset[ADS/Sa.CXO#Obs/11479]{11479} & 55234 &  2 &    &$-11.62\pm{0.03}$ & $2.3\pm{0.1}$
\enddata
\tablecomments{We provide Obsid, MJD, and exposure time of each
  \emph{Chandra} observation in ksec along with 2--10\,keV flux, and
  photon index ($\Gamma$) with their 1$\sigma$ uncertainties.  When
  only an upper limit could be determined, we list the $3\sigma$ upper
  limit.  Values of $\Gamma$ without uncertainties were fixed at 1.7.
  Sources GH 106, GH 119, GH 146, and GH 203 are archival data sets
  originally analyzed by \citet{2012ApJ...761...73D},
  \citet{2014ApJ...782...55Y}, \citet{2007ApJ...656...84G}, and
  \citet{2012ApJ...761...73D}, respectively.  Values for source 47 are
  from \emph{XMM-Newton} observations with results due to
  \citet{reneeinprep}.}
\kgtabend{xrayobs}{X-ray observations}

\kgfigbeg{fedd}
\includegraphics[width=0.95\columnwidth]{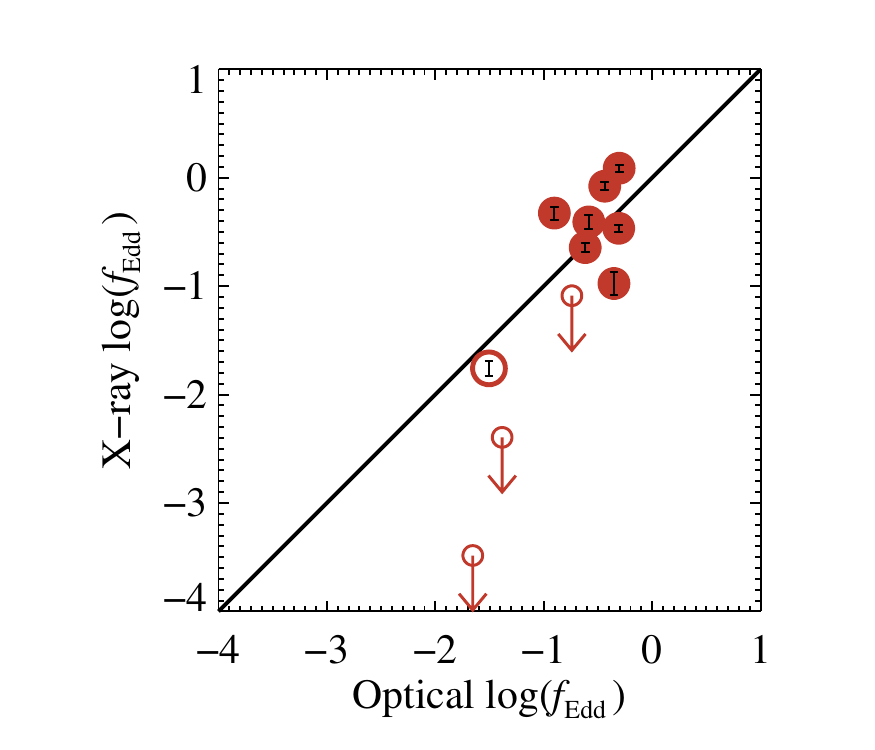}
\caption{Comparison of $f_\mathrm{Edd}$ estimates.  The abscissa shows
  $f_\mathrm{Edd}$ determined by \citet{gh07} from optical spectra.
  The ordinate shows $f_\mathrm{Edd}$ determined from 2--10 keV
  luminosities and bolometric corrections due to
  \citet{2004MNRAS.351..169M}.  Since the X-ray bolometric corrections
  depend on the bolometric luminosity, we use $L_\mathrm{bol}$ as
  determined by \citet{gh07}, which serves as a check for
  self-consistency.  Open circles indicate c-sample sources.  The
  error bars come from the uncertainty on the X-ray flux only, and the
  line shows equality.  There is generally good agreement, and the
  points above $f_\mathrm{Edd}=1$ are consistent with sub-Eddington
  accretion when uncertainties in black hole mass and bolometric
  correction are taken into account.}

\kgfigend{fedd}{Eddington fraction}

\subsection{Radio Observations}
\label{radioobs}

Radio data presented in this Letter come from new VLA observations,
taken at 8.4 GHz with 2 GHz bandwidth in the A configuration.  The
data were from two programs targeting low-mass AGNs, one selecting
from those with VLA FIRST detections, one from those without.
All observations began with a scan on a flux calibrator.  3C 286 was 
used for all observations, except GH 57 and GH 69, for which 3C 147 
was used.  The flux
calibrator was followed by a phase calibrator (Table
\ref{t:radioobs}).  Then we cycled between the source and the phase calibrator
for the remaining duration of the scan.  

Flagging and reduction of VLA data followed the standard pipeline
using CASA version 4.0.1.  We calibrated  fluxes based on the most
recent calibration models, implemented phase corrections, and then
calibrated the bandpass.  We averaged the data in bins in time by 30 s
and in frequency by eight channels.  Each frequency was converted into an
image and processed with the CLEAN algorithm separately with a maximum
of 5000 iterations, a gain of 0.1, and natural weighting.  The
processing used the full width of the $512\times512$ image at a
resolution of 0\farcs05 and typically stopped at a threshold of 0.05
mJy.

The resulting processed images were then inspected for emission at the
location of the AGN.  In 10 of the 12 sources, there was an
unambiguous, unresolved point source at the coordinates of the galaxy
center. We attribute this flux to the AGN.  For these detected
sources, we calculated the flux density by fitting a two-dimensional
Gaussian to the point source in a $20\times20$ pixel box and using the
total flux reported by the imfit tool.  Uncertainties in flux were
calculated as the quadrature sum of uncertainty in the fit and rms
noise of the image.  All detections were highly significant
($>10\sigma$).  The final flux densities we report (Table
\ref{t:radioobs}) are from the channel centered at 8.5 GHz with an
additional 5\% uncertainty to account for absolute flux calibration.
Non-detections are reported as $3\sigma$ upper limits of the rms
noise of the image.

Detection of the sources in the VLA FIRST survey was not a good
predictor of AGN radio emission.  The two non-detections are detected
in VLA FIRST, indicating that the 1.4 GHz VLA FIRST detection is not
associated with AGN activity.  All sources without detection at 1.4
GHz were detected at 8.5 GHz.  For the detected sources, we attempted
to constrain the spectral index, $\alpha$ ($S_\nu\propto
\nu^{-\alpha}$), by fitting a power law to the fluxes across the
entire bandpass of the radio observations, but the relatively large
uncertainties meant that we only had a weak constraint.  All
measurements were consistent with $\alpha=0.7$, which we assume for
subsequent analysis.

\kgtabstarbeg{rlrlllr@{\extracolsep{0pt}}ll}
  \footnotesize
  \tablecaption{VLA Observations}
  \tablewidth{\textwidth}
  \tablehead{
     \colhead{GH ID} &
     \colhead{SB ID} &
     \colhead{MJD} &
     \colhead{Cal.} &
     \colhead{$t_\mathrm{exp}$} &
     \colhead{Beam Size} &
     \multicolumn{2}{c}{$S_\nu/\mathrm{mJy}$} &
     \colhead{$\mathrm{rms}/\mathrm{mJy}$}
  }
  \startdata
 47 & 12097202 & 56298 &   J0830+2410 & 59.5 & $0\farcs37\times0\farcs28$ &    &$0.93  \pm{0.05}$ & 0.014 \\
 69 & 12452868 & 56274 &   J0914+0245 & 34.7 & $0\farcs36\times0\farcs26$ &    &$0.60  \pm{0.03}$ & 0.015\\
 87 & 12465022 & 56274 &   J1008+0730 & 33.2 & $0\farcs42\times0\farcs26$ &    &$0.42  \pm{0.03}$ & 0.022\\
 94 & 12469924 & 56274 &   J0958+6533 & 33.9 & $0\farcs32\times0\farcs25$ & $<$&$0.29$ & 0.087\\
101 & 12118330 & 56243 &   J1033+6051 & 57.6 & $0\farcs33\times0\farcs25$ &    &$0.54  \pm{0.03}$ & 0.012\\
106 & 12117646 & 56246 &   J1110+6028 & 57.6 & $0\farcs31\times0\farcs25$ &    &$1.10  \pm{0.06}$ & 0.014\\
140 & 12465449 & 56276 &   J1217+5835 & 33.8 & $0\farcs28\times0\farcs24$ &    &$0.37  \pm{0.03}$ & 0.018\\
146 & 11470757 & 56252 &   J1229+0203 & 56.3 & $0\farcs42\times0\farcs26$ &    &$0.45  \pm{0.03}$ & 0.017\\
158 & 12465639 & 56276 & J1256$-$0547 & 30.9 & $0\farcs30\times0\farcs24$ &    &$0.57  \pm{0.03}$ & 0.023\\
163 & 12465813 & 56277 &   J1347+1217 & 32.1 & $0\farcs49\times0\farcs26$ &    &$0.50  \pm{0.03}$ & 0.022\\
174 & 12469748 & 56276 &   J1419+5423 & 34.1 & $0\farcs37\times0\farcs24$ & $<$&$0.10$ & 0.028\\
203 & 12118194 & 56225 &   J1602+3326 & 57.4 & $0\farcs27\times0\farcs26$ &    &$0.58  \pm{0.04}$ & 0.015
\enddata
\tablecomments{VLA scheduling block identification, MJD
  of observation, gain calibrator, time on source in minutes, size of synthesized beam, flux density of source, and rms of map in mJy.  All observations used 3C286 for
  flux calibration and bandpass calibration except for GH 47 and GH
  69, which used 3C147.  The flux density uncertainty includes a
  5\% uncertainty in absolute flux calibration, which dominates the
  total uncertainty.  Upper limits are $3\sigma$
  values.}  
\kgtabstarend{radioobs}{VLA observations}

\section{Analysis and Discussion}
\label{analysis}
\label{discussion}

Our primary analysis is to compare these low-mass AGNs with the earlier
sample of SMBHs with dynamical masses on the fundamental plane.  As
mentioned in Section \ref{sample}, we update the \citet{gh07} masses
using the most recent H$\alpha$ AGN-mass scaling relations and adopt
conservative uncertainties of 0.5 dex.  The uncertainty in mass is
much larger than the statistical uncertainties in the H$\alpha$
luminosity and line-width measurements, the uncertainties in the
best-fit AGN-mass scaling relations, and the intrinsic scatter in the
scaling relations.  The larger uncertainty, however, should encompass
systematic uncertainties in mass estimation such as extrapolation of
scaling relations to low masses \citep[e.g.,][]{2010ApJ...721...26G},
linking H$\beta$ scaling relations to H$\alpha$ scaling relations, and
imperfect decomposition of narrow lines and broad lines, which may be
especially difficult for narrow-line Seyfert 1 AGNs
\citep{2009ApJ...692..246D}.

We calculate 5 GHz radio luminosities from our measured 8.5 GHz flux
densities assuming a spectral index of $\alpha=0.7$.  Note that $L_R\equiv\nu L_\nu$, whereas the 2--10 keV X-ray luminosity, $L_X$, is
a bandpass luminosity.

From the 13 sources in Table \ref{t:sample}, we have usable data in
both X-ray and radio for 10, including 3 sources with upper limits on
one of the two quantities.  We compare these data to the two
fundamental-plane fits in Figure \ref{f:fpview} and find that they are
better predicted by the universal (all black holes) fit.  The low-mass
AGNs are inconsistent with the SMBH-only relation, having higher $L_R$,
lower $L_X$, and/or lower $M$ than predicted.  Compared to the
universal relation, the low-mass AGNs are within the scatter.

\kgfigstarbeg{fpview}
\includegraphics[width=0.45\textwidth]{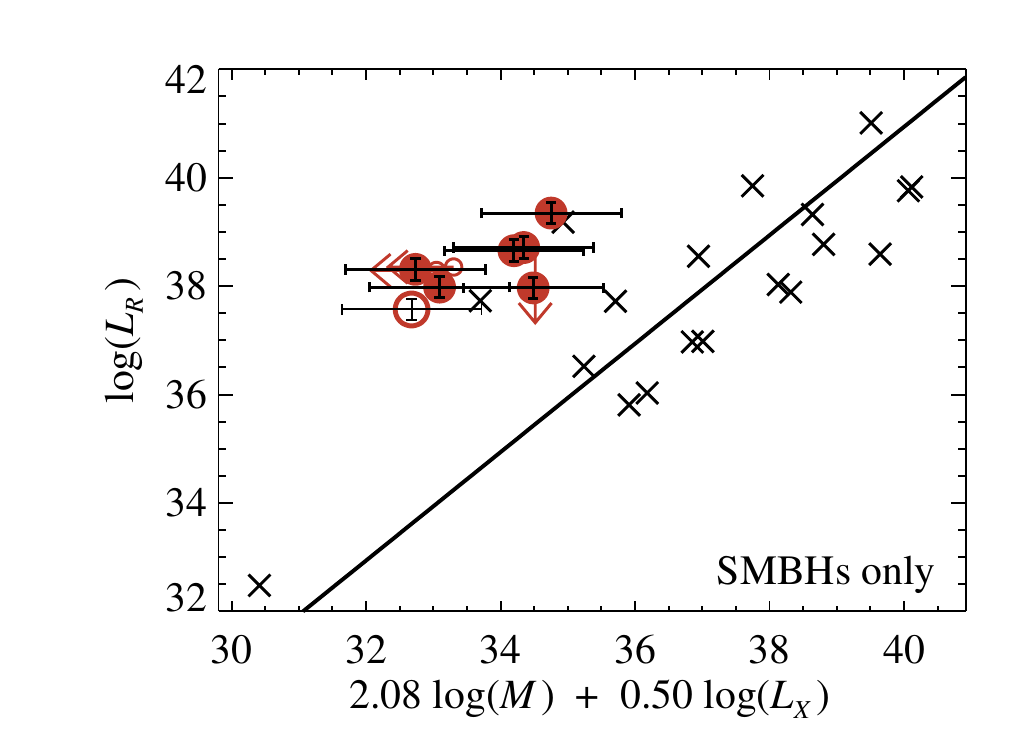}
\includegraphics[width=0.45\textwidth]{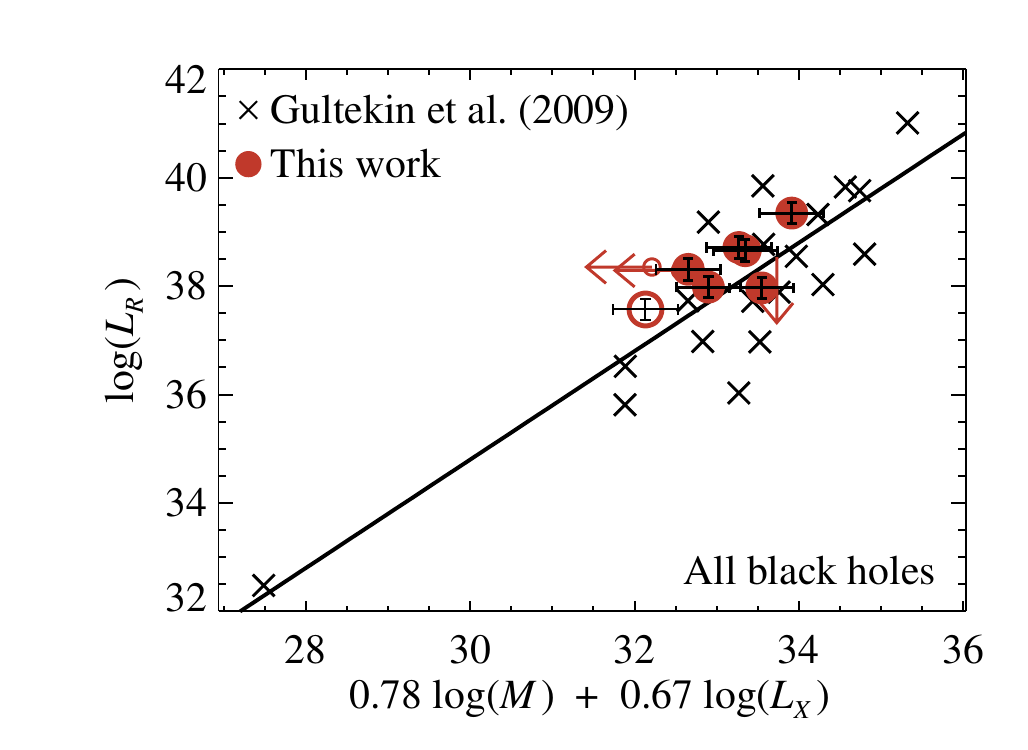}\\
\includegraphics[width=0.45\textwidth]{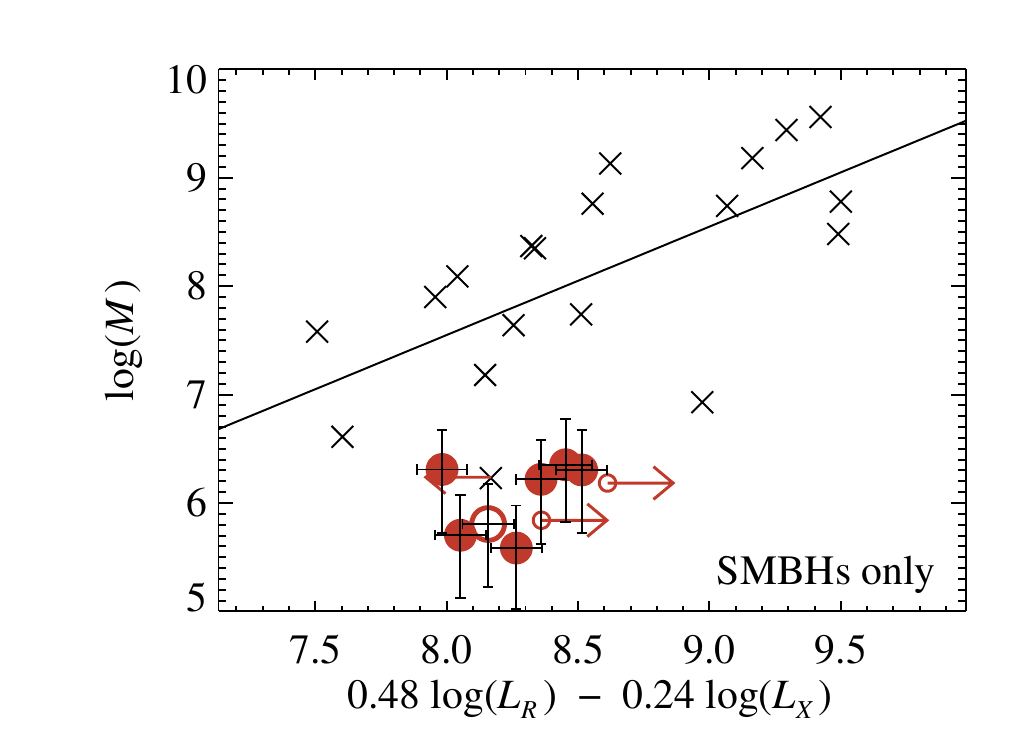}
\includegraphics[width=0.45\textwidth]{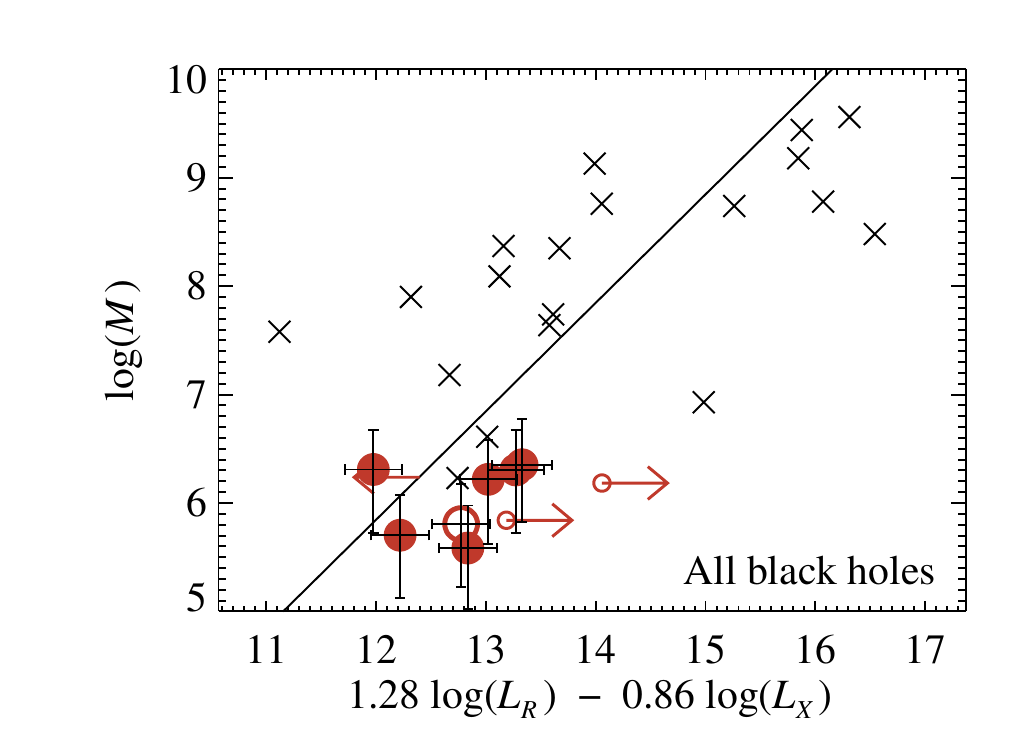}

\caption{New data for the current sample of small AGNs (red circles) and
  for SMBHs with dynamically determined black hole masses (black
  crosses) due to \citet{2009ApJ...706..404G}.  Open circles indicate
  sources from the \citet{gh07} c-sample.  Top panels have $L_R$ as
  the dependent variable, and bottom panels have $M$ as the dependent
  variable.  The left panels are projected to display the edge-on view
  of the best fit to SMBHs with dynamical masses only, and the right
  panels for both stellar-mass and SMBHs with
  dynamical masses.  Fits do not include low-mass AGNs, and thus we may compare the new, low-mass data to the values predicted by the fitted relations.  Low-mass AGNs
  are inconsistent with the prediction of the left panels and
  consistent with the prediction of the right panels.}

\kgfigstarend{fpview}{Fundamental plane  projections}

The low-mass AGNs are slightly but systematically offset from the
better-fitting SMBH-only plane in one direction (Figure \ref{f:fpview},
right panel).  The median deviations in the $L_R$, $M$, and $L_X$
directions are 0.63, $-0.81$, and $-0.95$ dex, respectively.  The
magnitude of these deviations is within the scatter, but of seven sources
with both X-ray and radio detections, six lie to one side of the plane.
If an individual source is equally likely to lie on either side of the
plane, the probability of six or more out of seven lying on the same side
(high or low, i.e., two-sided) by chance is 0.125.  If we
exclude the c-sample source GH 163 from this calculation, then chance
alignment for five of six sources would happen 0.219 of the time.  This is
only weak evidence of a systematic deviation, but we briefly
speculate on potential reasons if this trend were to continue with
more data.  Despite this speculation, below we conclude that the
low-mass AGNs are consistent with the full fundamental plane, which we
argue should be considered the better model.

Because black hole mass is a notoriously difficult quantity to
estimate and is the only indirect quantity considered, we give it
special attention.  The \citet{gh07} sample is selected for broad
H$\alpha$ lines with small FWHM.  It is
possible, but not certain, that broad emission lines come from the
base of a disk wind \citep[e.g.,][]{2013A&A...551L...6K} with
predominantly rotational motions.  The resulting FWHM measurement
would be subject to orientation effects that could lead to a
preferential selection of face-on inclinations and underestimation of
the true mass.  This possibility is supported by the LAMP 2008 campaign's
finding that lower inclination objects return larger virial
coefficients \citep{2013arXiv1311.6475P}.

Several other lines of evidence, however, point to, on average,
accurate mass estimation.  First, reverberation mapping of GH 126
indicates a small mass in agreement with the H$\alpha$ single-epoch
estimate \citep{2011ApJ...741...66R}.  Second, the expected black hole
mass based on host galaxy stellar velocity dispersion generally agrees
with the single-epoch estimates \citep{2011ApJ...739...28X}, though
megamaser measurements of black hole mass in small galaxies show that
the masses can be much smaller than predicted by velocity dispersion
\citep{2010ApJ...721...26G} and low-mass AGNs appear undermassive
relative to their bulge luminosity \citep{2011ApJ...737L..45J}.  Given
that many of the host galaxies are likely to be pseudobulges
\citep{2011ApJ...742...68J}, black hole mass--host galaxy property
scaling relations may not apply to the majority of these galaxies
\citep{2011Natur.469..374K, 2010ApJ...721...26G}.  Finally,
\citet{reneeinprep} recently found that the X-ray variability of a
different sample of \citet{gh07} AGNs is as expected for low-mass
sources.  Given the above evidence, we find it unlikely that
underestimated masses are the cause of the offset.

Another possibility is that the offset results from different
accretion--jet interactions.  The data are not conclusive, but X-ray
properties of AGNs may depend on black hole mass
\citep{2009ApJ...698.1515D, 2012ApJ...761...73D}.  Given the inferred
near-ultraviolet luminosity of these low-mass AGNs, their spectral
index between 2500 \AA\ and 2 keV ($\alpha_{OX}$) may be flatter than
expected, and their 2 keV X-ray emission sometimes weaker than
expected, though selection effects may complicate the issue.  This may
be evidence of slim-disk accretion, lack of a Comptonizing corona, or
high intrinsic absorption.  Thus the 2 keV X-ray emission of low-mass
AGNs may not be as reliable a probe of accretion power as it is in
higher mass AGNs.  If the difference were a result of a lack of a
Comptonizing corona, then this would lead to an absence of 2--10 keV
flux, which we assume is from inverse Compton scattering.  Rather, we
observe X-ray flux levels that are consistent with expectations based
on the accretion rate and bolometric corrections (Figure \ref{f:fedd}).
If the difference in 2 keV luminosity is, instead, a result of high
intrinsic absorption of any form, our conclusions should be unaffected
owing to our use of hard flux and deeper observations.

Based on the relatively small offset, the reliability of 2--10 keV
flux measurements in the face of absorption, and the low significance
of a possible offset of the low-mass points from the fundamental
plane, we conclude that low-mass AGNs do in fact follow the full
$M$--$L_R$--$L_X$ plane.  Our results suggest that stellar-mass black
holes and supermassive black holes follow the same relation.  A
similar test using the most massive black holes would provide
additional support.

Given the low mass and relatively high accretion rates here, we can
now state with confidence that (1) it is appropriate to use the
fundamental plane to estimate SMBH masses smaller
than $\sim10^7\ \msun$ \citep[e.g.,][]{2011Natur.470...66R}, (2) it
is possible to use the fundamental plane to test for intermediate mass
black holes \citep[e.g.,][]{2013MNRAS.436.1546M}, and (3) it is
possible to use the fundamental plane to estimate black hole masses at
high accretion rates \citep[e.g.,][]{2011ApJ...738L..13M}.

\hypertarget{ackbkmk}{}%
\acknowledgements 
\bookmark[level=0,dest=ackbkmk]{Acknowledgments}

We thank Jenny Greene for a valuable referee report and Elena Gallo
for helpful comments.  K.G.\ acknowledges support provided by NASA
through Chandra Award G02-13111X (\emph{Chandra X-Ray Observatory} operated by NASA under contract
NAS8-03060).  Results reported here are based on new and archival
\emph{Chandra} data.  We used NED, ADS, and CIAO and are grateful.

\bibliographystyle{apjads}
\hypertarget{refbkmk}{}%
\bookmark[level=0,dest=refbkmk]{References}
\bibliography{gultekin}

\label{lastpage}
\end{document}